\documentclass[a4paper,twoside]{article}
\usepackage[authoryear]{natbib}
\bibpunct{(}{)}{;}{a}{}{,}
\usepackage{graphicx}
\date{} %Please leave the date blank
\title{\large\bf\flushleft The Metallicity Distribution of Distant F and G type stars in the CFHT Legacy Survey Deep Field}
\author{\parbox{\textwidth}{\flushleft
\vspace{-0.5cm}
{\it Y\"uksel Karata\c{s}$^1$, Mukremin Kilic$^2$, Orhan  G\"{u}ne\c{s}$^1$, F\"usun Limboz$^1$}\\
\vspace{0.4cm}
{\small $^1$Istanbul University, Science Faculty, Department of Astronomy 
and Space Sciences, 34119, Istanbul, T\"{u}rkiye\\
$^2$Department of Astronomy, The Ohio State University, Columbus, OH 43210, USA}\\
{\small Email: karatas@istanbul.edu.tr}}}
\begin{document}
\twocolumn[
\begin{changemargin}{.8cm}{.5cm}
\begin{minipage}{.9\textwidth}
\vspace{-1cm}
\maketitle
%
%
%%%%%%%%%%%%%     ABSTRACT    %%%%%%%%%%%%%
%Abstract of no more than 200 words here.
\small{\bf Abstract}: 
We study the metal abundances of F and G type stars in the Galactic disk
and halo using the Canada$-$France$-$Hawaii Telescope Legacy Survey D4 field.
For a sample of stars within 7 kpc of the Galactic plane, we derive mean
abundance values of \lbrack Fe/H\rbrack $=-0.77\pm0.36$ dex for the thick disk, and 
\lbrack Fe/H\rbrack $=-1.42\pm0.98$ dex for the stellar halo, respectively. 
These metallicites are consistent with the
metallicity estimates from the Sloan Digital Sky Survey Data Release 6. 
We do not find a vertical abundance gradient for the thick disk between 1 kpc and 
4 kpc. However, stars within 1 kpc of the Galactic plane are saturated in 
our dataset, and therefore we cannot rule out the observed vertical metallicity 
gradient for the thick disk stars in the $SDSS$. Moreover, we find a negative
trend in the halo metallicity with increasing distance from the Galactic plane.
This trend could be due to a contribution from an increasing number of 
very metal poor stars that belong to the outer halo. However, systematic 
effects in photometric metallicities are largest for the most
metal poor stars, and therefore these effects cannot be ruled out.

%%%%%%%%%%%%%     KEYWORDS    %%%%%%%%%%%%%
\medskip{\bf Keywords:} stars: abundances: stars: distances: techniques: photometric 
% Please write all keywords in lower case. PASA uses the
% standard list of subject headings adopted by The Astrophysical Journal
% and available from http://www.journals.uchicago.edu/ApJ/keywords_text.html.
% Keywords are separated by em-dashes, i.e. ---

%%%%%%%%DO NOT EDIT%%%%%%%%%%%%
\medskip
\medskip
\end{minipage}
\end{changemargin}
]
\small
%%%%%%%%EDIT FROM HERE%%%%%%%%%%%%

\section{Introduction}

Our view of the formation of the Milky Way has changed dramatically since
the discoveries of complex substructures in the Galaxy. The Milky Way
has four spatially and kinematically distinct components; thin disk, 
thick disk, halo, and the bulge.
However, recent discoveries of irregular 
structures like the Sagitarrius dwarf tidal stream and the Monoceros 
stream show that the formation of galaxies is not a steady process
resulting in a smooth distribution of stars. Instead, galaxies are 
constantly shaped by the infalling smaller galaxies (see Juric et al. 2008
and the references there in). 

The Sloan Digital Sky Survey ($SDSS$) and the Sloan Extension for Galactic 
Understanding and Exploration ($SEGUE$) are providing photometry and 
spectroscopy for a very large number of stars that can be used to study 
the stellar populations in the Galaxy. Such a study involving $\approx20,000$
stars with the SDSS spectra demonstrated that the Galactic halo
is likely to have two components; an inner and an outer component 
(Carollo et al. 2007). These two components seem to have different spatial
density profiles, stellar orbits, and metallicities. \cite{Carolloetal2007} 
suggest that the outer halo formed through dissipationless chaotic
merging of smaller subsystems within a pre-existing dark matter halo, 
supporting the complex galaxy formation scenarios. \cite{Ivezicetal2008}
use the SDSS photometry data to study the metallicity distribution of 
disk and halo stars. However, their dataset is not deep enough to test 
the dichotomy of the halo. 

The Canada$-$France$-$Hawaii Telescope Legacy Survey ($CFHTLS$)
provides a valuable new source to study the metallicity distribution of
the Galactic stellar populations. The third data release of the Deep survey
now provides $ugriz$ photometry down to a limiting magnitude of 26.
$CFHTLS$ lack spectroscopic observations, which would provide the most 
accurate metallicities. However, imaging data can also be used to deduce 
stellar metallicities through the traditional UV excess method 
(Wallerstein 1962; Sandage 1969). The metal absorption lines mostly affect
the UV part of the spectrum, therefore metal$-$rich and metal$-$poor stars with
the same effective temperature (or the same $B-V$ color) can be differentiated
based on their UV colors. In addition, imaging data enables us to obtain a 
complete flux-limited sample of stars and push the studies of stellar 
metallicities to fainter magnitudes than the spectroscopic
surveys. Here we use the $CFHTLS$ data for the D4 Field to perform such an
analysis. Section~2 describes the data and the selection of point sources. 
The derivation of metallicites from the photometric data and its application
to the $CFHTLS$ $ugriz$ system is discussed in Section~3, whereas the results 
from this analysis are discussed in Section~4. 

\section {The Data}

The CFHTLS D4 field is located at RA = $22^h 15^m 31^s$, DEC = $-17^{o} 
43^{'} 56^{''}$ (J2000), ${\it l}=39^{o}$, and ${\it b}=-53^{o}$. The TERAPIX T003
public data release for the D4 field covers an area of 0.86 deg$^2$ and 
provides photometry with limiting AB magnitudes of $u=$ 26.5, $g=$ 26.3, 
$r=$ 26.4, $i=$ 26.0, and $z=$ 25.0, respectively. The high galactic latitude
of this field is similar to the fields targeted by the $SDSS$.

Without follow-up spectroscopy, we have to rely on morphological classification
to select point sources. \cite{Schultheisetal2006} and \cite{Limbozetal2008} 
showed that the star$-$galaxy separation is possible through
the use of the half$-$light radius (hereafter, HLR) measurements from the 
$i$-band images. HLR corresponds to the radius that encloses 50\% of the 
total flux from the source. Figure 1 shows the HLR versus $i$-band magnitudes
for the sources in the D4 field. Objects with $i < 17$ mag are saturated, and
therefore have unreliable HLR measurements. Using spectroscopic data from 
the VLT$-$VIRMOS Deep Survey ($VVDS$; Le Fevre et al. 2005), \cite{Limbozetal2008}  
demonstrated that the spectroscopically confirmed stars have HLR($i)<2.82$ 
and that the star$-$galaxy separation is reliable down to $i=21$ mag.
Since the $CFHTLS$ goes five magnitudes deeper than this, we now look into 
pushing the point source classification limit to fainter magnitudes.

\begin{figure}[t!]
\resizebox{\hsize}{!}
{\includegraphics[]{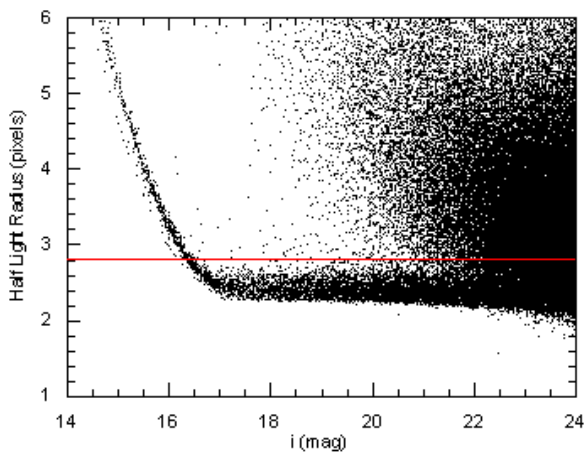}}
\caption{HLR versus $i$-band for the D4 field.
Note that objects with $i < 17.00$ are saturated.
Star$-$galaxy separation seems to be robust down to $i$ = 22 mag.
The solid horizontal line marks the boundary for selection of
point sources ($HLR < 2.82$).}
\label{figure1}
\end{figure}

Figure~2 presents $(u-g)$ versus $(g-r)$ color$-$color diagrams for four
different magnitude bins. The top left panel shows the color$-$color diagram
for point sources (black points) and resolved sources (red points) for 
$i=20-21$ mag, where the point source classification seems reliable. 
This panel reveals a tight stellar sequence, and confirms the expectations 
that stars can be identified reliably down to 21 mag. The other panels in 
this figure show the color$-$color diagrams for $i=21-22, 22-23$ and $23-24$ mag.
These three panels show that the stellar sequence is still clearly visible 
for $i=21-22$, but the galaxy contamination starts to become a problem 
for $i>22$ mag. We count the number of sources that are classified as 
point-like and that lie to the left of the stellar sequence,
and compare it with the number distribution of resolved sources with 
HLR($i)> 2.82$ to estimate contamination of the stellar sequence by galaxies.
We estimate that the galaxy contamination is less than 4\% in the range 
$(g-r)=0.1-0.7$ mag for $i=21-22$ mag. An analysis of the number distribution
of resolved and un-resolved sources in the other panels show that
the galaxy contamination ranges from 7\% to 23\% for $i=22-23$ mag, and 
even higher for $i>23$ mag.  Therefore, we restric our sample to the 
sources with $i < 22$ mag  and $HLR < 2.82$. 
Using the above criteria, we classify 7348 sources as stars.

\begin{figure}[b!]
\centering
%\special{bmp: f02.bmp x=6cm y=6cm}
\includegraphics[scale=0.4, angle=0]{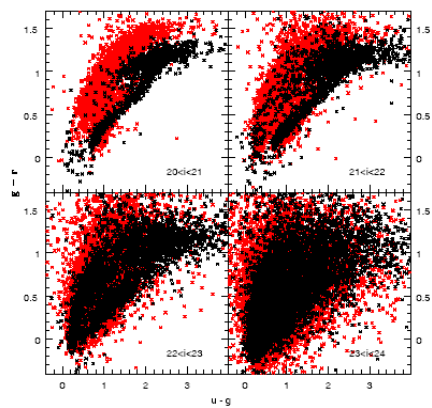}
\caption{$(u-g)$ versus $(g-r)$ color$-$color diagram for different
$i$-filter magnitude limits. Black points shows stars ($HLR < 2.82$),
whereas red points represent galaxies ($HLR \ge 2.82$).}
\end{figure}

We use the dust maps of \cite{Schlegeletal1998} to de-redden the photometry.
E($B-V$) for the D4 field ranges from 0.023 to 0.031 mag, with a mean of 0.027 mag.
The small range of reddening correction for the D4 field shows that this correction
does not have a signifincant effect on our analysis.
Perhaps the most important correction required for the photometry is the
correction from the $MegaCam$ system to the standard $ugriz$ filter system.
Studying the differences between the $ugriz$ system and the $MegaCam$ system,
\cite{Clemetal2008} find that the $MegaCam$ $griz$ filters are consistent with 
the standard system, though the $u$ band data need to be corrected using 
a third order polynomial.   
The $MegaCam$ $u-$band filter was selected to take advantage of the better 
UV transparency of the $CFHT$ and it is slightly redder than the $SDSS$ 
$u-$ band filter. \cite{Clemetal2008} compare theoretical isochrones in the standard 
system and the corrected $MegaCam$ system and they find that the isochrones 
overlap each other within 0.01 mag for both dwarfs and giants with colors 
$u-g$ = 0.0 to +3.0 mag and $g-i=-0.5$ to +3.5 mag, and with metallicities 
[Fe/H] = $-2.3$ to 0.0. We use the transformation given by \cite{Clemetal2008} 
(see their Figure 5) to convert the $MegaCam$ $u-$band photometry into the 
standard $u-$band system. In addition, a correction is required to transform
the photometry from the primed system to the $SDSS$ 2.5 m telescope's unprimed 
$ugriz$ system. We use the transformations given by \cite{Tuckeretal2006} to 
transform the photometry to the unprimed system.

The apparent $g$ magnitude distribution for our sample of stars is shown in 
Figure 3a. The number of stars increases with increasing magnitude until 
$g=$ 22.1 mag, which we use as the completeness limit. For a $\approx$ G5 
main$-$sequence star with $(g-r)$ = 0.4 mag (see Table 1 in Ivezic et al. 2008),
this limit corresponds to a Galactocentric distance of 21 kpc
and a Galactic plane distance of $\approx$17 kpc.
Figure~3b shows the $(g-r)$ color distribution for our sample.
There are two distinct peaks located at $\approx$ 0.20 mag and 1.15 mag, 
corresponding to F stars and late type M stars, respectively. 
We select our sample of F and G type stars by restricting the sample to 
the stars with $0.20 < (g-r) < 0.60$ (shown by vertical dashed lines). 
This results in a sample of 1315 F and G stars. Majority of these stars lie 
above the synthetic $(u-g)$ vs. $(g-r)$ relation of \cite{Pickles1998}, 
indicating that they are metal$-$poor disk and halo stars.

\section {The Metallicity Distribution}

Using $SDSS$ spectroscopy and photometry, \cite{Ivezicetal2008} have developed
a metallicity estimation method based on the $(u-g)$ and $(g-r)$ colors.
Their method reproduces the metallicities obtained from low resolution $SDSS$
spectroscopy with a root-mean-square scatter of 0.2 dex.
They also derive a photometric parallax relation using $ugriz$ photometry of 
globular clusters. The agreement between the $CFHT/MegaCaM$ and the $SDSS$ 
$ugriz$ filter sets, after correcting the $MegaCam$ $u-$band photometry 
(Clem et al. 2008), enables us to use their photometric metallicity and parallax
methods on our dataset. Photometric metal abundances of the F and G
dwarfs with $0.2<(g-r)\leq0.4$, are determined via the following equation of \cite{Ivezicetal2008}. 

\begin{eqnarray}
[Fe/H]=- 4.37-8.56(u-g)+15.5(g-r)\nonumber\\
-39.0(u-g)(g-r)+23.5(u-g)^2+20.5(g-r)^2\nonumber\\
+12.1(u-g)^2(g-r)+7.33(u-g)(g-r)^2\nonumber\\
-10.1(u-g)^3-21.4(g-r)^3  
\end{eqnarray}

For G dwarfs with $(g-r) > 0.4$, we replaced the $(u-g)$ color in the above equation
with $(u-g)-2(g-r)+0.80$. We estimate the absolute magnitudes of our sample of stars
using the following relations from \cite{Ivezicetal2008}.

\begin{eqnarray}
M_{r}(g-i,[Fe/H])=M_{r}(g-i)+\Delta M_{r}([Fe/H])
\end{eqnarray}
\begin{eqnarray}
\Delta M_{r}([Fe/H]) = 4.50-1.11[Fe/H]-0.18[Fe/H]^2 
\end{eqnarray}  
\begin{eqnarray}                                                                    	
M_{r}(g-i)=-5.06+14.32(g-i)-12.97(g-i)^2\nonumber\\
+6.127(g-i)^3-1.267(g-i)^4+0.0967(g-i)^5	
\end{eqnarray}                                               

\begin{figure} [t!]
\centering
\resizebox{\hsize}{!}
{\includegraphics[]{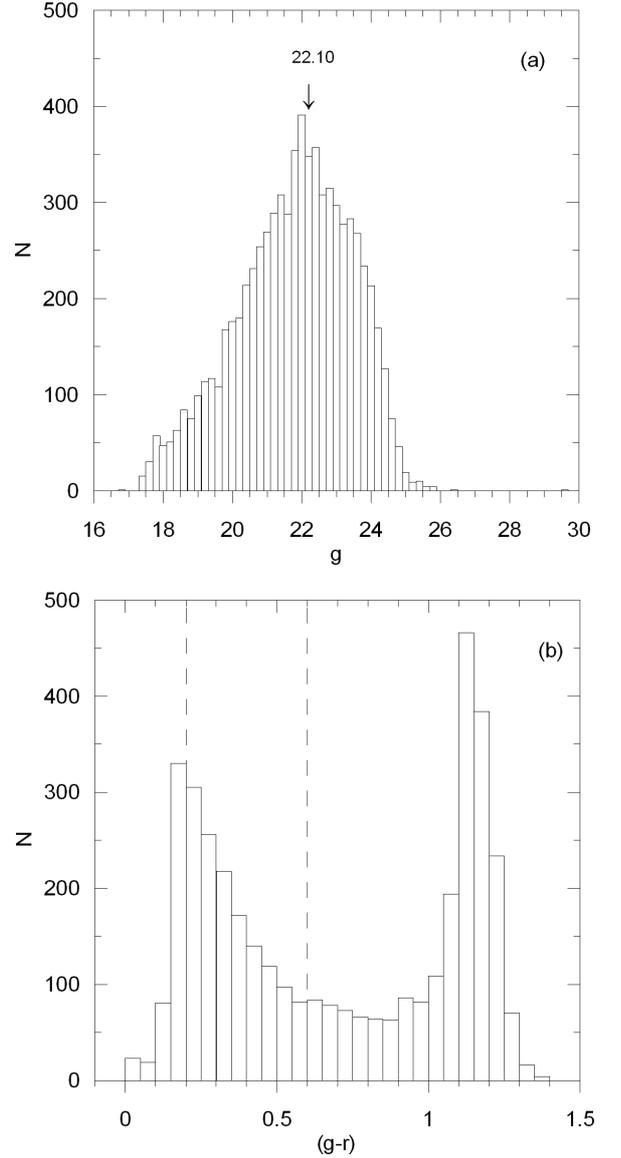}}
\caption{(a) $g$ magnitude distribution. Drop in $g= 22.10$
is taken as the completeness limit. (b) $(g-r)$ distribution.
The presence of the two main peaks at $(g-r)\sim0.20$ and
$\sim1.15$ are seen. Vertical dashed lines show the color index range
$0.20 < (g-r) < 0.60$, which corresponds to F and G type stars.}
\end{figure}

\begin{figure} [h!]
\centering
\resizebox{\hsize}{!}
{\includegraphics[]{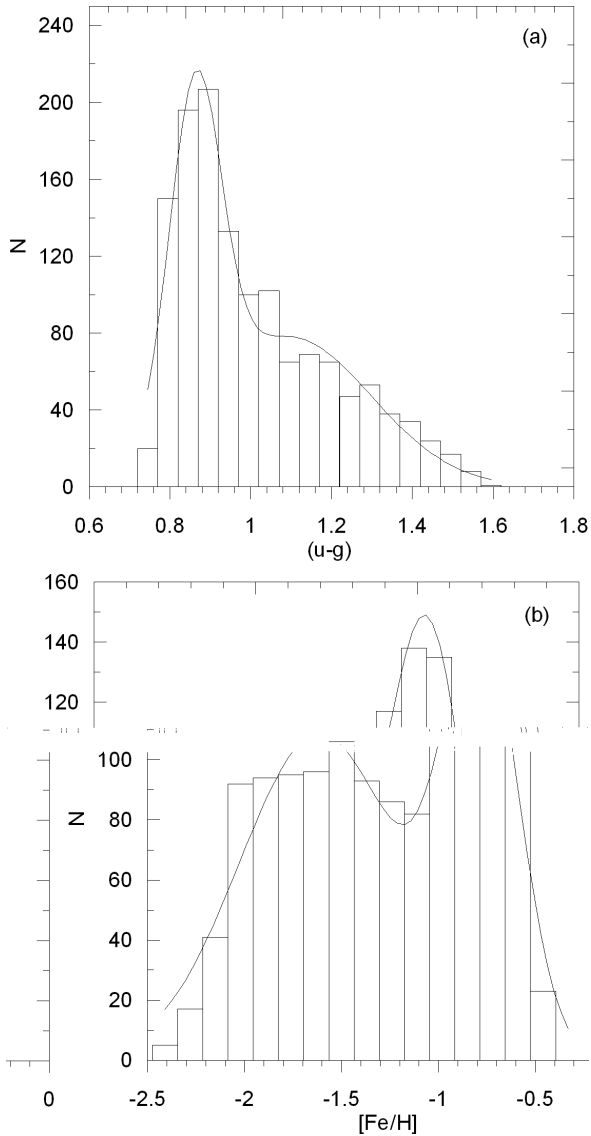}}
\caption{(a) $(u-g)$ color and (b) [Fe/H] distribution for our sample of 1315 F and G
type stars. Solid lines represent two Gaussian distributions fitted to the data.}
\label{figure4}
\end{figure}

\begin{figure}[h]
\includegraphics[scale=0.45, angle=0]{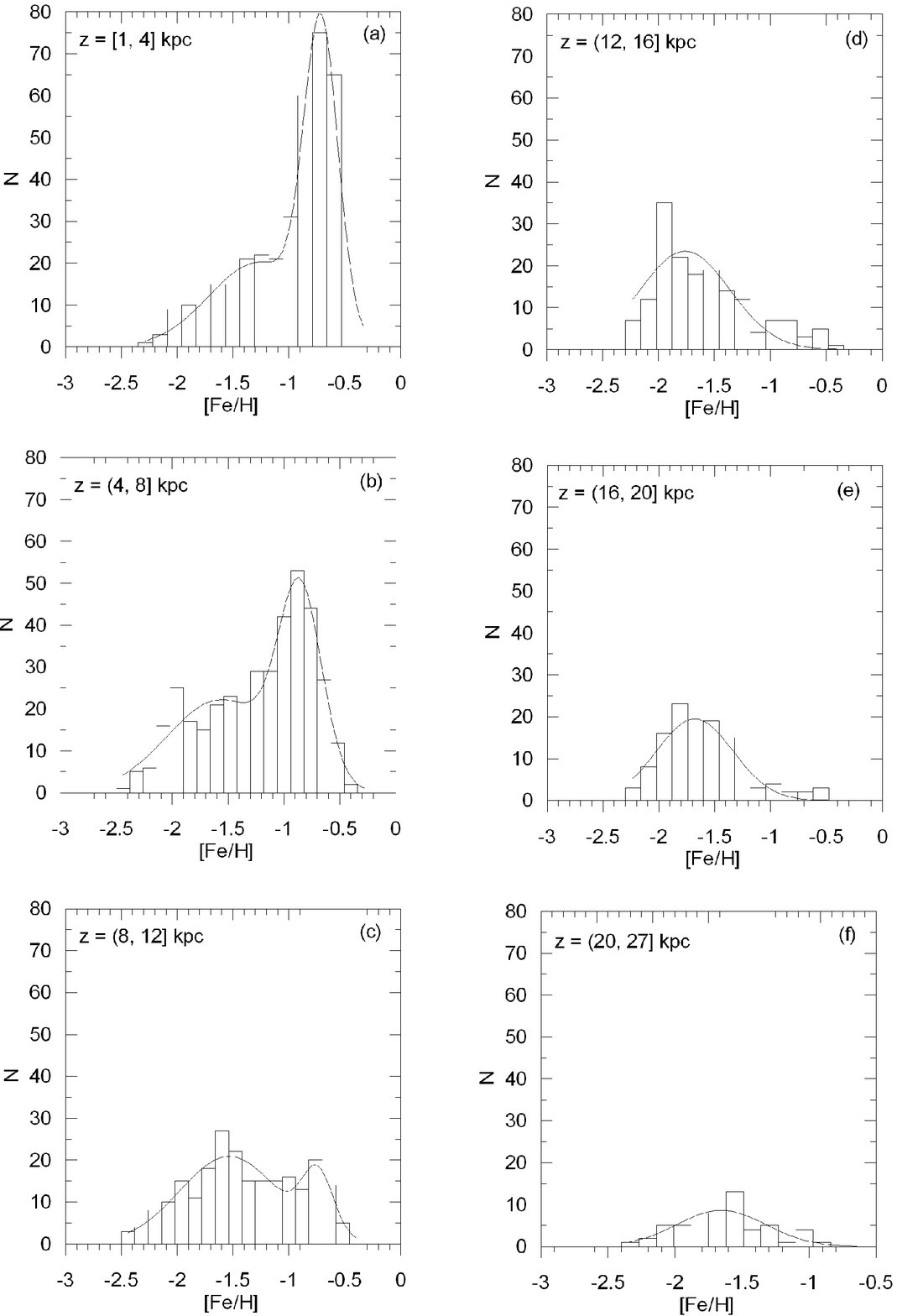}
\caption{The Metallicity distribution as a function of the distance from the plane of the Galaxy.}
\label{figexample}
\end{figure}

Figure 4 displays the $u-g$ color distribution and the derived metallicities for
our sample of F and G stars. Both of these distributions are bimodal, and they are
fairly well fitted with two gaussians. \cite{Ivezicetal2008} also found a bimodal 
$u-g$ color distribution for the $SDSS$ sample with two peaks at $u-g\approx 0.9$  
and 1.1 mag. The two peaks observed in the color distribution of our sample are 
located at $u-g= 0.86 \pm 0.13$ and $1.10 \pm 0.41$ mag. The agreement between 
the colors of our sample of F and G stars and that of the $SDSS$ F and G
stars shows that the transformations between the $MegaCam$ photometry and the 
standard $ugriz$ system are reliable. The two peaks in the $u-g$ color distribution
correspond to metal$-$poor halo stars and more metal$-$rich and
closer disk stars, respectively. The width of 0.13 mag is significantly larger
than the median $u-g$ error of 0.02 mag at $g=22$ mag, providing a measure of the
intrinsic variations in the $u-g$ color and the metallicity distribution.

The observed peaks in the $u-g$ color distribution translates into two peaks
at [Fe/H] = $-0.78 \pm 0.39$ and $-1.62 \pm 0.82$ in the metallicity distribution.
Figure 5 shows this distribution as a function of increasing distance from the
Galactic plane. The mean metal abundances for the fitted gaussians are also 
listed in Table 1. The mean metal abundance shifts as a function of galactic 
plane distance. The thick disk dominates at $z = 1-4$ kpc, and a small contribution
is still visible up to $8-12$ kpc, after which only the halo stars with [Fe/H] 
$\approx -1.7$ are visible.

\section{Discussion}

Our mean metallicity estimate of [Fe/H] = $-0.78 \pm 0.39$ for the thick disk is
consistent with the results from the analysis of F and G stars in the $SDSS$.
Using the $SDSS$ Data Release 3 spectroscopy, \cite{AllendePrieto2006} found 
that thick disk G dwarfs with $1<z<3$ kpc show a compact metallicity distribution
with a peak at [Fe/H] $\approx -0.70$. Our mean metallicity estimates for
$z$ = 1$-$2, 2$-$3, and 3$-$4 kpc (see Table 1) range from $-0.71$ to $-0.73$ dex, and 
they are entirely consistent with the Allende Prieto et al. result. However, 
the mean metallicity for the thick disk is more metal$-$poor compared to the 
literature values of $-0.60$ \citep{GilmoreWyse1985} and $-0.50$ \citep{Carneyetal1989}.
Since the $SDSS$ photometric metallicity relation is known to underestimate 
the metallicity at high metallicities (see the discussion in Ivezic et al. 2008), 
our mean metallicity estimate for the thick disk probably suffers
from these systematic effects.

Using the $SDSS$ Data Release 6 photometry, \cite{Ivezicetal2008} found that the 
median metallicity for the thick disk smoothly decreases with distance from 
the plane from [Fe/H] = $-0.60$ at 500 pc to $-0.80$ at several kpc. We do not 
observe such a metallicity gradient for the thick disk between $z$ = 1 kpc and 
4 kpc, but our mean metallicity estimate is consistent with their mean metallicity
estimate beyond several kpc, $-0.80$ dex. In fact, if we restrict our sample to 
$z<7$ kpc, then the mean metallicity peak for the thick disk stars in our sample
shifts to [Fe/H] = $-0.77 \pm 0.36$, entirely consistent with the Ivezic et al. analysis.
Since stars with $i<17$ mag are saturated in the $CFHTLS$ data, we do not have any
stars within 1 kpc of the Galactic plane. The trend found by Ivezic et al. is 
mostly based on the relatively brighter stars in their sample\footnote{This sample may be
contaminated by metal-rich stars that belong to the high velocity tail of the thin disk distribution.}, 
which show a peak in metallicity at [Fe/H] $\approx -0.60$. The majority of the $z>1$ kpc thick disk 
stars in their sample show a peak around [Fe/H] $\approx -0.70$ (see their Figure 7), 
similar to our analysis. 

\begin{table*} [t]
\begin{center}
\tiny
\caption{Metal abundances for the thick disk and halo
stars as a function of distance from the Galactic plane.}
\small
\begin{tabular}{lccccc}
\hline
 $z$(kpc) &$<z>$ & $[Fe/H]_{\rm THICK}$ & $\sigma_{[Fe/H]}$ & $[Fe/H]_{\rm HALO}$
& $\sigma_{[Fe/H]}$ \\
\hline
$[1.0,~2.0]$   &1.5   & $-$0.73 & 0.30 & -  & - \\
$(2.0,~3.0]$   &2.5   & $-$0.73 & 0.36 & $-$1.36  & 1.02\\
$(3.0,~4.0]$   &3.5   & $-$0.71 & 0.29 & $-$1.36  & 0.88\\
\hline
$[1.0,~4.0]$   &2.5   & $-$0.71 & 0.29 & $-$1.27  & 0.89\\
$(4.0,~8.0]$   &6.0   & $-$0.85 & 0.38 & $-$1.57  & 0.96\\
$(8.0,~12.0]$  &10.0  & $-$0.72 & 0.28 & $-$1.54  & 0.95\\
$(12.0,~16.0]$ &14.0  & -       & -    & $-$1.77  & 0.69 \\
$(16.0,~20.0]$ &18.5  & -       &      & $-$1.67  & 0.68\\
$(20.0,~27.0]$ &23.5  & -       &  -   & $-$1.65  & 0.62\\
\hline
\end{tabular}
\end{center}
\end{table*}

Figure 6 shows the metallicity trends for the thick disk and halo as a function 
of the Galactic plane distance. If we ignore the thick disk metallicity estimates
beyond 8 kpc due to small number statistics, then a small shift in metallicity
is observed for the thick disk between 4 and 8 kpc. However, the mean metallicities
for the stars with $z$ = 1$-$4 kpc and 4$-$8 kpc are consistent within the errors.
In addition, we do not find any significant trend between the Galactocentric distance
and metallicity for the thick disk stars.

Our metallicity estimate of $-1.62 \pm 0.82$ for the halo is consistent with 
the literature values of $-$1.60 dex \citep{RyanNorris1991} and $-$1.70 dex 
\citep{Carneyetal1996}. However, it is more metal poor than the halo metallicity 
derived using the $SDSS$ F and G stars (Allende Prieto et al. 2006;
Ivezic et al. 2008). Allende Prieto et al. (2006) found that the halo stars exhibit
a broad range of iron abundances with a peak at [Fe/H] = $-1.40$, whereas 
\cite{Ivezicetal2008} find a peak at $-$1.46 dex with a standard deviation of 0.30 dex.
Restricting our sample to $z <7$ kpc, as in the Ivezic et al. study, we find that
the mean metallicity for the halo stars within $z$ = 7 kpc is [Fe/H] = $-1.42$. 
Therefore, our metallicity estimates for both the thick disk and halo are 
consistent with the results from the $SDSS$ analysis of stars within $z$ = 7 kpc.

\begin{figure} [t!]
\centering
\resizebox{\hsize}{!}
{\includegraphics[]{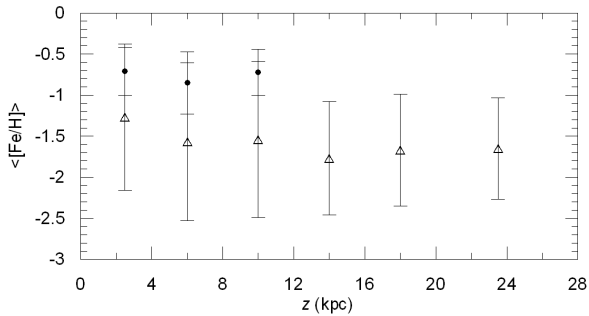}}
\caption{The Mean metallicity for the thick disk (filled dots) and
halo (open triangles) as a function of distance from the Galactic plane.}
\end{figure}

Since our results are consistent with the results from the $SDSS$, and the 
$CFHTLS$ data goes deeper than the $SDSS$, we can use the $CFHTLS$ data to study 
the stars more than 7 kpc away from the Galactic plane.  Our dataset is complete
up to $z$ = 17 kpc for $g-r = 0.4$ stars and up to 11 kpc for the entire sample 
(see Section 2). Figure 6 shows that the mean metallicity for the halo decreases with
increasing distance from the plane of the Galaxy. 
However, the metallicities are also consistent within the errors.
If the decline in halo metallicity is real, this trend in metallicity
and the observed broad distribution may be due to
the inclusion of an increasing number of very metal poor stars in the halo.

Using spectroscopic and kinematic data from the $SDSS$ Data Release 5, \cite{Carolloetal2007}
found a dichotomy in the metallicity distribution of the halo stars with two peaks at
[Fe/H] $\approx -1.6$ and [Fe/H] $< -2.0$. They found that stars that reach large 
distances in their orbits above the Galactic plane show a metallicity
peak around $-$2.1 dex, and they identified those stars as members of the outer halo. 
Since the number of these outer halo stars with retrograde orbits is small 
compared to the inner halo stars, it is impossible to identify them without 
kinematic information. However, a comparison of the observed metallicity 
distribution for our sample of stars and the fitted double gaussians in Figure 5
shows that the metal$-$poor tail end of the gaussian distribution for the halo 
underestimates the number of observed stars for $z>4$ kpc. The majority of the
panels in Figure 5 show a nearly flat distribution of stars between [Fe/H] = $-1.70$ and $-2.0$.
This could be caused by a contribution from the outer halo stars. However, without 
kinematic information, we cannot confirm or reject this idea. In addition, 
the systematic effects for the photometric metallicity estimates are largest
for the most metal$-$poor stars. \cite{Ivezicetal2008} find that the root-mean-square 
scatter of the residuals between the photometric and spectroscopic metallicity 
estimates of the F and G stars in the $SDSS$ increases up to 0.30 dex for low 
metallicity stars. Even though we corrected for the differences between the
$MegaCam$ $u$ filter and the standard filter set, a small error in
this transformation may also result in systematic offsets. For example, an 
offset of 0.04 mag in the $u-$band data can cause a metallicity bias of up to
0.20 dex (Ivezic et al. 2008). Hence, the observed distribution may be 
affected by these systematic problems.

\section{Conclusions}

We have performed a morphological selection of point sources and analyzed the 
metallicity distribution of distant F and G type stars in the $CFHTLS$ D4 field.
We have used $MegaCam$ photometry transformed to the standard $ugriz$ filter system,
and used the photometric parallax and metallicity relations from the $SDSS$
to obtain metallicities and distances for our sample of 1315 F and G type stars.  

Limiting our study to $z$ = 7 kpc, the distance limit used by \cite{Ivezicetal2008},
we find that our estimates for the mean metallicity of the thick disk 
([Fe/H] = $-0.77$) and halo ($-$1.42 dex) are consistent with the results from
the $SDSS$ Data Release 6 photometry ($-$0.80 dex and $-$1.46 dex). Since the $CFHTLS$ data
is deeper than the $SDSS$, we are also able to check for trends in the thick disk
and halo metallicity distribution beyond 7 kpc. Unlike the Ivezic et al. study,
we do not find a trend in metallicity for the thick disk between 1 and 4 kpc. 
However, the significance of the trend observed in the $SDSS$ dataset comes from 
the stars within 1 kpc of the Galactic plane. Unfortunately, those stars are 
saturated in the $CFHTLS$ observations, and the metallicity trend for the thick 
disk stars further away than 1 kpc from the plane is consistent with
the $SDSS$ results. 

We observe a decline in metallicity of the halo with increasing distance from 
the Galactic plane. We suggest that this decline may be due to an increasing 
contribution from the very metal poor ([Fe/H]$<-2$) halo stars, but we cannot 
rule out systematic problems in our analysis. Proper motion data from the 
$CFHTLS$ observations will be available at the end of the survey period. 
These data will be useful to study the kinematic properties of our sample
of stars and constrain the contribution of very metal poor stars. 

\section{Acknowledgements}

We thank D. An for useful discussions, C. Benoist for providing us with
the CFHTLS data, and the anonymous referee for useful comments. This work is partly based on observations obtained with
MegaPrime / MegaCam, a joint project
of CFHT and CEA/DAPNIA, at the $Canada-France-Hawaii$ Telescope (CFHT) 
which is operated by the National Research Council (NRC) of Canada, 
the Institut National des Science de l'Univers of the Centre National
de la Recherche Scientifique (CNRS) of France, and the University of 
Hawaii. This work is based in part on data products produced at TERAPIX 
and the Canadian Astronomy Data Centre as part of the Canada-France-Hawaii
Telescope Legacy Survey, a collaborative project of NRC and CNRS.

\end{document}